\documentstyle[12pt]{article}

\begin{document} 

\begin{titlepage}

\hrule 
\leftline{}
\leftline{Preprint
          \hfill   \hbox{gr-qc/9905024}}
\leftline{\hfill   \hbox{April 1999}}
\leftline{\hfill   \hbox{revised August 1999}}
\vskip 5pt
\hrule 
\vskip 1.0cm

\centerline{\large\bf 
Energy Conservation and the Unruh Effect 
} 

\vskip 1cm

\centerline{{\bf 
Riuji Mochizuki$^{1}{}^{\dagger}$
and Takayuki Suga$^2{}^{\ddagger}$
}}  
\begin{description}
\item[]{\it  
$^1$ Laboratory of Physics, Tokyo Dental College\\
1-2-2 Masago, Mihama-ku, Chiba 261-8502,  Japan
  }
\item[]{\it 
$^2$ Graduate School of Science and Technology, Chiba University\\
Yayoi, Inage-ku Chiba 263-8522, Japan
  }
\item[]{$^\dagger$ 
  E-mail:  rjmochi@tdc.ac.jp 
  }
\item[]{$^\ddagger$ 
  E-mail:  psuga@cuphd.nd.chiba-u.ac.jp   
  }
\end{description}

\centerline{{\bf Abstract}} \vskip .5cm

In this paper it is explicitly demonstrated that the energy 
conservation law is kept when a detector uniformly accelerated in 
the Minkowski vacuum is excited and emits a particle.  This fact 
had been hidden in conventional approaches in which detectors were 
considered to be forced on trajectories.    To lift the veil we
suggest a detector model written in terms of the Minkowski 
coordinates.  In this model the Hamiltonian of the detector 
involves a classical potential term instead of the detector's 
fixed trajectory.  The transition rate agrees with the corresponding 
conventional one in the limit of an infinite mass detector though 
even then the recoil remains.

\vskip 0.5cm

\hrule  

\vskip 0.2cm  


\end{titlepage}


\section{Introduction}

It is more than a quarter of a century that has passed since his 
vital paper was given by Fulling~\cite{full}.  He pointed out that 
an inertial observer and a uniformly accelerated one would 
construct the operator algebras which are not unitarily equivalent 
to each other.  The Bogoliubov transformation between them 
demonstrates a well-known effect which is often called the Unruh 
effect~\cite{unru,davi}.  If the Minkowski vacuum is described 
by a uniformly accelerated observer, it is not a state in which 
there are no particles but a thermal bath (the so-called 
Fulling-Davies-Unruh thermal bath) which is characterized by a 
temperature proportional to his proper acceleration.  This 
observation carries a serious question of the particle concept 
and necessity of preparing measuring devices. DeWitt~\cite{dewi} 
provided a point-like detector model with two internal energy 
levels which is coupled linearly to a scalar field.  Using this 
model, the Unruh effect consequentially contains a paradoxical 
event: a DeWitt detector uniformly accelerated in the Minkowski 
vacuum can be excited and emit the radiation.  Many authors seeked 
a source of this mysterious energy.  For example, Unruh and 
Wald~\cite{unwa} and Takagi~\cite{taka} casted their consideration 
over vacuum fluctuation of the field.  Birrell and 
Davies~\cite{bida} suspected that extra work done by the external 
force which accelerated the detector was the source of this energy.  
In their approach, however, the detector's trajectory is fixed.  
This curtains energy and momentum conservations.  Though 
Parentani~\cite{pare} treated the detector's trajectory as a 
dynamical variable, a description of the flow on energy was not 
given.

The main purpose of this paper is to clarify the fact that the 
energy of the radiation and difference of the detector's rest mass 
comes directly from the kinetic energy via the recoil of the 
detector.  In other words, proved in this paper is that the Unruh 
effect is a phenomenon satisfying the energy conservation law.  To 
this end, the conventional DeWitt detector model is modified into 
a form in which the detector is not fixed on any trajectories.  
Instead, the Hamiltonian of the detector involves a classical 
potential term if it is considered to be accelerated.  The rest 
mass of the detector treated as finite throughout calculation, the 
transition rate obtained using this model agrees with 
conventional one after taking the limit of the infinite mass 
detector.  In our calculation it is essential that the 
translational invariance is broken due to the potential energy of 
the detector.  

This paper is organized as follows.  In the second section the 
transition rate of a detector in inertial motion is discussed 
to demonstrate difference between this case and  that of the 
accelerated detector.  In the third section the transition 
rate corresponding to the Fulling-Davies-Unruh thermal bath 
is recreated as that of the detector moving in classical potential 
with a constant gradient.  In the last section we summarize the 
result and make a remark.  
\section{Detector model with no fixed trajectory}

A detector which is coupled with a scalar field via a monopole 
interaction is called a DeWitt detector. In the conventional 
DeWitt detector model, the detector is supposed to move along a 
trajectory $x^{\mu}(\tau)$, where $\tau$ is the detector's proper 
time.   The interaction Hamiltonian $H_I$ is
\begin{equation}
H_I=\int d^3{\bf x}\int^{\infty}_{-\infty}d\tau c_0m(\tau)\phi(x)
\delta^{(4)}(x-x(\tau)),
\end{equation} 
where $c_0$ is a coupling constant and $\phi$ is a scalar field 
which interacts with the detector.  The detector's monopole moment 
operator $m(\tau)$ is written in the interaction picture as
\begin{equation}
m(\tau)={\rm e}^{iH_0\tau}m(0){\rm e}^{-iH_0\tau},
\end{equation}
\begin{equation}
m(0)=\mid m><m_0 \mid ,
\end{equation}
where $m$ and $m_0$ are the rest masses on the detector's upper and 
lower energy levels, respectively.  The free Hamiltonian $H_0$ 
is defined on the detector's trajectory, so that 
\begin{equation}
H_0\mid m> =m \mid m>.
\end{equation}

The amplitude for the transition in which a detector is excited 
and simultaneously emits a scalar particle with its momentum {\bf k} 
in Minkowski vacuum $\mid 0_M>$ is given by first order 
perturbation theory as
\begin{eqnarray}
A&=& -i<m,1_{\bf k}\mid \int d^4x\int^{\infty}_{-\infty}
d\tau c_0m(\tau)\phi(x)\delta^{(4)}(x-x(\tau))
\mid 0_M,m_0>\nonumber \\
&\ & -i<m,1_{\bf k}\mid\int^{\infty}_{-\infty}c_0m(\tau)\phi[x(\tau)]
d\tau\mid 0_M,m_0>.
\end{eqnarray}
$\phi$ is expanded in terms of Minkowski plane wave modes, thus
\begin{equation}
<1_{\bf k}\mid\phi(x)\mid 0_M>={n\over \sqrt{\omega}}
{\rm e}^{i\omega t-i{\bf k}\cdot{\bf x}},
\end{equation}
where $n$ is a normalization constant and $ \omega^2=\mid{\bf k}\mid^2$
 for the real particle.  
Substituting (2.2), (2.3), (2.4) and (2.6) into (2.5) yields
\begin{equation}
A= -i{nc_0\over\sqrt{\omega}}\int^{\infty}_{-\infty}d\tau{\rm e}
^{i\Delta m\tau
+i(\omega t(\tau)-{\bf k}\cdot{\bf x}(\tau))}, 
\end{equation}
where 
\[
\Delta m=m-m_0.
\]

To discuss physical significance of the different vacua and 
energy conservation of the Unruh effect, detector models should 
be examined carefully.   In the conventional DeWitt model, as 
noted above, a trajectory of the detector is imposed and the 
coordinates in the transition amplitudes are treated as functions 
of the detector's proper time.  This is a dynamically rather misty 
manipulation.  Hence we suggest a modified DeWitt detector model 
in which the detector moves in classical potential but is not forced 
on any trajectories. 

First, we consider the case in which the detector moves inertially.  
We introduce the detector fields $\Phi_0$ and $\Phi$ with rest masses
$m_0$ and $m$, respetively.  They satisfy the Klein-Gordon equation
\[
(\overline\sqcup+m_0^{\ 2})\Phi_0(x)=0,
\]
\[
(\overline\sqcup+m^2)\Phi(x)=0
\]
and are expanded in complete sets of solutions of the above 
equations:
\begin{eqnarray}
&\Phi_0&(x)=\Phi_0^+(x)+\Phi_0^-(x) \\
& &\Phi_0^{+}(x)=\int d^3{\bf P}_0{N_0\over \sqrt{E_0}}a_0({\bf P}_0)
{\rm e}^{-i(E_0t-{\bf P}_0\cdot{\bf x})} \\
& &\Phi_0^{-}(x)=\int d^3{\bf P}_0{N_0\over \sqrt{E_0}} a_0^{\ \dagger}
({\bf P}_0){\rm e}
^{+i(E_0t-{\bf P}_0\cdot{\bf x})},
\end{eqnarray}
\begin{eqnarray}
&\Phi&(x)=\Phi^+(x)+\Phi^-(x) \\
& &\Phi^{+}(x)=\int d^3{\bf P}{N\over \sqrt{E}}a({\bf P}){\rm e}
^{-i(Et-{\bf P}\cdot{\bf x})} \\
& &\Phi^{-}(x)=\int d^3{\bf P}{N\over \sqrt{E}} a^{\dagger}
({\bf P}){\rm e}
^{+i(Et-{\bf P}\cdot{\bf x})},
\end{eqnarray}
\begin{equation}
E_0=({\bf P}_0^{\ 2}+m_0^{\ 2})^{1/2},\ \ 
E=({\bf P}^2+m^2)^{1/2},
\end{equation}
where $N_0$ and $N$ are normalization constants and $a_0$ ($a_0^{\ \dagger}$) 
and $a$ ($a^{\dagger}$) are annihilation (creation) operators of the 
detector with rest masses $m_0$ and $m$, respectively.

In terms of these fields, the interaction Hamiltonian density ${\cal H}_{\rm I}$ is defined as
\begin{equation}
{\cal H}_{\rm I}=c\Phi^-(x)\Phi^+_0(x)\phi(x),
\end{equation}
where $c$ is a coupling constant.  The transition amplitude corresponding to
(2.5) is
\begin{equation}
A=-i<{\bf P},m\mid<1_{\bf k}\mid c\int d^4x\Phi^-(x)\Phi^+_0(x)\phi(x)
\mid 0_M>\mid m_0,{\bf P}_0>,
\end{equation}
where
\[
\mid m,{\bf P}>=a^{\dagger}({\bf P})\mid 0>,
\]
\[
\mid m_0,{\bf P}_0>=a_0^{\ \dagger}({\bf P}_0)\mid 0>.
\]
Using the expansions (2.9) and (2.13), this amplitude becomes
\begin{eqnarray}
A&\sim&{1\over\sqrt{EE_0\omega}}\int d^4x{\rm e}^{i(E-E_0+\omega)t
-i({\bf P}-{\bf P}_0+{\bf k})\cdot {\bf x}}\\
&\sim&{1\over\sqrt{EE_0\omega}}\delta(E-E_0+\omega)
\delta^{(3)}({\bf P}-{\bf P}_0+{\bf k}).
\end{eqnarray}

These delta functions explicitly indicate energy and momentum 
conservations at the interaction between the detector and the 
field $\phi$.  The amplitude (2.18) always vanishes because the 
arguments of these delta functions cannot be zero 
simultaneously~\cite{suga}.  Indeed, when the momentum 
conservation is used, the argument of the first delta function 
becomes
\begin{eqnarray}
E-E_0+\omega&=& m\gamma -m_0\gamma_0+\omega\nonumber \\
&=& \gamma_0^{\ -1}(m\tilde\gamma-m_0)+(m\gamma{\bf V}-m_0
\gamma_0{\bf V}_0)\cdot{\bf V}_0+\omega\nonumber \\
&=& \gamma_0^{\ -1}(m\tilde\gamma-m_0)+(\omega -{\bf k}\cdot
{\bf V}_0),
\end{eqnarray}
where ${\bf V}_0={\bf V}_0({\bf P}_0)$ and ${\bf V}={\bf V}({\bf P})$ 
are velocities of the detector before and after the interaction, 
respectively, and the $\gamma$ factors are defined as
\[
\gamma_0=(1-\mid{\bf V}_0\mid^2)^{-1/2},
\]
\[
\gamma=(1-\mid{\bf V}\mid^2)^{-1/2},
\]
\[
\tilde\gamma=(1-\mid\tilde{\bf V}\mid^2)^{-1/2}=\gamma\gamma_0
(1-{\bf V}\cdot{\bf V}_0),@\ \ (\tilde{\bf V}:{\rm relative}
\ {\rm velocity}).
\]
Recalling $m>m_0$, $\tilde\gamma >1$ and $\omega>{\bf k}\cdot
{\bf V}_0$, the right-hand side of (2.19) is always positive.  To 
argue its correspondence with the conventional DeWitt model, we go 
back to the amplitude (2.17).  Perfoming the {\bf x}  
integration and using (2.19), this becomes
\begin{eqnarray}
A&\sim&{1\over\sqrt{EE_0\omega}}\int dt
{\rm e}^{i(m\gamma({\bf P}_0,{\bf k}) -m_0\gamma_0({\bf P}_0)+\omega)t}
\delta^{(3)}({\bf P}-{\bf P}_0+{\bf k})
\nonumber \\
&\sim&{1\over\sqrt{EE_0\omega}}\int dt
{\rm e}^{i(\gamma_0^{\ -1}(m\tilde\gamma-m_0)+(\omega -{\bf k}
\cdot{\bf V}_0))t}
\delta^{(3)}({\bf P}-{\bf P}_0+{\bf k}).
\end{eqnarray}

In the limit of an infinite mass detector($m\rightarrow\infty$, $m_0
\rightarrow\infty$ while maintaining $\Delta m$ finite,
\[
{\bf V}_0={\bf V},\ \ \tilde{\bf V}=0.
\]
Then (2.20) becomes
\begin{equation}
A\sim{1\over \sqrt{\omega}}\int d\tau
{\rm e}^{i\Delta m\tau+i(\omega t(\tau)-{\bf k}\cdot{\bf x}(\tau))}
\delta^{(3)}({\bf P}-{\bf P}_0+{\bf k}),
\end{equation}
where $\tau$ is the detector's proper time, 
and $t(\tau)$ and ${\bf x}(\tau)$ are defined as
\[
t(\tau)=\gamma_0\tau,
\]
\[
{\bf x}(\tau)={\bf V}_0t(\tau).
\]
What is considered is the transition in which the rest mass of the
detector undergoes the change $m_0\ \rightarrow\ m$ irrespective of 
the final momentum of the detector.  Hence the transition rate $R$ is
\begin{eqnarray}
R&\sim&\int d^3{\bf P}\int d^3{\bf k}
\mid A\mid^2\nonumber \\
&\sim& \int d^3{\bf P}\int d^3{\bf k}{1\over \omega}
\int d\overline{\tau}\int d(\Delta \tau)
{\rm e}^{i\Delta m\Delta\tau+i(\omega t(\Delta\tau)-{\bf k}
\cdot{\bf x}(\Delta\tau))}
\Big[ \delta^{(3)}({\bf P}-{\bf P}_0+{\bf k})\Big]^2\nonumber \\
&\sim&\int d^3{\bf k}{1\over \omega}
\int d\overline{\tau}\int d(\Delta \tau)
{\rm e}^{i\Delta m\Delta\tau+i(\omega t(\Delta\tau)-{\bf k}
\cdot{\bf x}(\Delta\tau))},
\end{eqnarray}
where
\[
\overline{\tau}={1\over 2}(\tau +\tau^{'}),
\]
\[
\Delta\tau=\tau -\tau^{'}
\]
and we have used
\[
\Big[\delta^{(3)}({\bf P}-{\bf P}_0+{\bf k})\Big]^2
\sim\delta^{(3)}({\bf P}-{\bf P}_0+{\bf k}).
\]
This agrees with that obtained using the conventional 
DeWitt model up to a constant factor.
Performing the $\Delta\tau$ integration, 
we obtain
\begin{equation}
R\sim\delta(\Delta m+\gamma_0 (\omega -{\bf k}\cdot {\bf V}_0))=0.
\end{equation}
This transition rate for the inertial detector always vanishes. 
Note that it is not 
due to energy conservation but due to the fact that energy and 
momentum conservations are never concomitant for $\Delta m\ge 0$.
\section{the Unruh effect}
\setcounter{equation}{0}

In this section we demonstrate the Unruh effect as response of a 
detector moving in a classical scalar potential $-Fz$ in which $F$ 
is a constant.  
The Klein-Gordon equations the detector fields $\Phi_0~{'}$ and 
$\Phi^{'}$ should satisfy are
\[
\Bigl[(i{\partial\over\partial t}+Fz)^2+\nabla^2-m_0^{\ 2}\Bigr]
\Phi_0^{'}=0,
\]
\[
\Bigl[(i{\partial\over\partial t}+Fz)^2+\nabla^2-m^{\ 2}\Bigr]
\Phi^{'}=0.
\] 
We can obtain the solutions of these equations by the aid of WKB
approximation method.  In the classically allowed region, the detector's
fields are expanded as
\begin{eqnarray}
\Phi_0^{'}(x)&=&\Phi_0^{'+}(x)+\Phi_0^{'-}(x) \\
\Phi_0^{'+}(x)&=&\int_0^{\infty}dE_0\int_{-\infty}^{\infty}dP_{0x}
\int_{-\infty}^{\infty}dP_{0y}{N_0^{'}\over \sqrt{\mid P_{0z}\mid}}a_0
(E_0,P_{0x},P_{0y})\nonumber \\
&&\times{\rm e}^{-i(E_0t-P_{0x}x-P_{0y}y-\int^zP_{0z}dz)}
\theta \Big(z-{1\over F}(E_0-\sqrt{m_0^{\ 2}+P_{0x}^{\ 2}+P_{0y}^{\ 2}}
)\Big) \\
\Phi_0^{'-}(x)&=&\int_0^{\infty}dE_0\int_{-\infty}^{\infty}dP_{0x}
\int_{-\infty}^{\infty}dP_{0y}{N_0^{'}\over \sqrt{\mid P_{0z}\mid}}
a_0^{\ \dagger}
(E_0,P_{0x},P_{0y})\nonumber \\
& &\times{\rm e}^{+i(E_0t-P_{0x}x-P_{0y}y-\int^zP_{0z}dz)}
\theta \Big(z-{1\over F}(E_0-\sqrt{m_0^{\ 2}+P_{0x}^{\ 2}+P_{0y}^{\ 2}}
)\Big),
\end{eqnarray}
\begin{eqnarray}
\Phi^{'}(x)&=&\Phi^{'+}(x)+\Phi^{'-}(x) \\
\Phi^{'+}(x)&=&\int_0^{\infty}dE\int_{-\infty}^{\infty}dP_{x}
\int_{-\infty}^{\infty}dP_{y}{N^{'}\over \sqrt{\mid P_{z}\mid}}a
(E,P_{x},P_{y})\nonumber \\
& &\times{\rm e}^{-i(Et-P_{x}x-P_{y}y-\int^zP_zdz)}
\theta \Big(z-{1\over F}(E-\sqrt{m^{2}+P_{x}^{\ 2}+P_{y}^{\ 2}}
)\Big) \\
\Phi^{'-}(x)&=&\int_0^{\infty}dE\int_{-\infty}^{\infty}dP_{x}
\int_{-\infty}^{\infty}dP_{y}{N^{'}\over \sqrt{\mid P_{z}\mid}}a^{\dagger}
(E,P_{x},P_{y})\nonumber \\
& &\times{\rm e}^{+i(Et-P_{x}x-P_{y}y-\int^zP_zdz)}
\theta \Big(z-{1\over F}(E-\sqrt{m^{2}+P_{x}^{\ 2}+P_{y}^{\ 2}}
)\Big),
\end{eqnarray}
where the mode functions have been selected to become plane wave when 
$F\rightarrow 0$ and
\begin{equation}
\mid P_{0z}\mid =\sqrt{(E+Fz)^2-m_0^{\ 2}-P_{0x}^{\ 2}-P_{0y}^{\ 2}},
\end{equation}
\begin{equation}
\mid P_{z}\mid =\sqrt{(E+Fz)^2-m^{2}-P_{x}^{2}-P_{y}^{2}}.
\end{equation}
The transition amplitude in this case is
\begin{equation}
A=-i<P_y,P_x,E,m\mid<1_{\bf k}\mid c\int d^4x\Phi^{'-}(x)\Phi_0^{'+}(x)
\phi(x)\mid 0_M>\mid m_0,E_0,P_{0x},P_{0y}>,
\end{equation}
where
\[
\mid m,E,P_x,P_y>=a^{\dagger}(E,P_x,P_y)\mid 0>,
\]
\[
\mid m_0,E_0,P_{0x},P_{0y}>=a_0^{\ \dagger}(E_0,P_{0x},P_{0y})\mid 0>.
\]
Using the expansions (3.2) and (3.6), this amplitude becomes
\begin{eqnarray}
A& \sim& \int d^4x{1\over\sqrt{\mid P_z\mid\mid P_{0z}\mid\omega}}
{\rm e}^{i(E-E_0+\omega)t-i(P_x-P_{0x}+k_x)x-i(P_y-P_{0y}+k_y)y}
{\rm e}^{-i(\int^zP_zdz-\int^zP_{0z}+k_zz)}\nonumber \\
& &\times
\theta \Big(z-{1\over F}(E-\sqrt{m^{2}+P_{x}^{\ 2}+P_{y}^{\ 2}})\Big)
\theta \Big(z-{1\over F}(E_0-\sqrt{m_0^{\ 2}+P_{0x}^{\ 2}+P_{0y}^{\ 2}}
)\Big)
\nonumber \\
& \sim& \int dz{1\over\sqrt{\mid P_z\mid\mid P_{0z}\mid\omega}}
{\rm e}^{-i(\int^zP_zdz-\int^zP_{0z}+k_zz)}\nonumber \\
&&\times\delta(E-E_0+\omega)\delta(P_x-P_{0x}+k_x)\delta(P_y-P_{0y}+k_y)
\nonumber \\
& & \times
\theta \Big(z-{1\over F}(E-\sqrt{m^{2}+P_{x}^{\ 2}+P_{y}^{\ 2}})\Big)
\theta \Big(z-{1\over F}(E_0-\sqrt{m_0^{\ 2}+P_{0x}^{\ 2}+P_{0y}^{\ 2}}
)\Big).
\end{eqnarray}
The first delta function in this equation explicitly indicates energy 
conservation, that is, the energy paid for the radiation comes from the 
detector's kinetic energy. 
This statement may give a somewhat peculiar impression if minding a 
fact that a slowly moving detector does not have enough kinetic 
energy to satisfy the above equation.  This is, however, only 
alarmism because there is no inertial frame in which the detector 
is always at rest and because the time when the interaction occurs 
cannot be exactly determined due to the uncertinly principle.

  To see correspondence with the conventional approach, we limit 
the detector's motion on $z$-axis:
\[
P_{0x}=P_{0y}=0.
\]
Then the transition amplitude (3.10) becomes
\begin{eqnarray}
A&\sim& \int dz{1\over\sqrt{\mid P_z\mid\mid P_{0z}\mid\omega}}
\delta (m\gamma-m_0\gamma_0+\omega)\delta (P_x+k_x)\delta(P_y+k_y)
{\rm e}^{-i\int^z(m\gamma V_z-m_0\gamma_0V_0+k_z)dz}\nonumber \\
& &\times
\theta \Big(z-{1\over F}(E-\sqrt{m^{2}+P_{x}^{\ 2}+P_{y}^{\ 2}})\Big)
\theta \Big(z-{1\over F}(E_0-m_0
)\Big),
\end{eqnarray}
where
\begin{eqnarray*}
\gamma&=&{E+Fz\over m}, \\
\gamma_{0}&=&{E_{0}+Fz\over m_{0}},\\
\mid V_{z}\mid &=&{\sqrt{(E+Fz)^2-m^{2}}\over E+Fz},\\
\mid V_{0z}\mid &=&{\sqrt{(E_0+Fz)^2-m_0^{\ 2}}\over E_0+Fz}.
\end{eqnarray*} 
Substituting (2.19) into the exponent in (3.11) yields
\[
m\gamma V_z-m_0\gamma_0V_{0z}+k_z=-{\omega\over V_{0z}}+k_z-
{m\tilde\gamma-m_0\over\gamma_0V_{0z}}.
\]

In the limit of the infinite mass detector, (3.11) becomes
\begin{eqnarray}
A\sim& & \int^{\infty}_{-{E_0-m_0\over F}}{dz\over\sqrt{\mid P_z\mid
\mid P_{0z}\mid \omega}}
\delta (m\gamma-m_0\gamma_0+\omega)\delta (P_x)\delta(P_y)\nonumber \\
& &\times\exp \Big[-i\int^z\Big(-{\omega\over V_{0z}}+k_z-
{m\tilde\gamma-m_0\over\gamma_0V_{0z}}\Big)dz\Big]\nonumber \\
\sim& & \int^{\infty}_{-\infty}{d\tau\over\sqrt{\omega}}
\delta (m\gamma-m_0\gamma_0+\omega)\delta (P_x)\delta(P_y)\nonumber \\
& &\times\exp\Big[+i\Big(\Delta m\tau+\omega t(\tau)-k_zz(\tau)
\Big)\Big],
\end{eqnarray}
where
\begin{eqnarray}
z(\tau)+{E_0\over F}&=&{m_0\over F}\cosh {F\tau\over m_0}, \\
t(\tau)&=&{m_0\over F}\sinh {F\tau\over m_0},
\end{eqnarray}
$\tau <0$ and $\tau >0$ correspond to $V_{0z}<0$ and $V_{0z}>0$, 
respectively.  The response rate in this case is~\cite{bida}
\begin{eqnarray}
R&\sim&\int dE\int dP_x\int dP_y\int d^3{\bf k}\mid A\mid^2\nonumber \\
&\sim&\int d^3{\bf k}{1\over \omega}\Bigl(\int d\tau
{\rm e}^{i\Delta\tau+i\omega 
t(\tau)-ik_zz(\tau)}\Bigr)^2\nonumber \\
&\sim&\int d\overline\tau {\Delta m\over {\rm e}^{2\pi\Delta m\cdot 
m_0/F}-1}.
\end{eqnarray}

  This result indicates that the detector moving in the 
classical potential $-Fz$ responds as if it were immersed in a 
thermal bath. This is known as the Unruh effect, that is,
the Unruh effect can be interpreted
as bremsstrahlung by a heavy particle with two internal energy level.
\section{summary and remark}

In this paper we have suggested a transcription of a DeWitt detector.  
In our approach the detector 
is not forced on any trajectories.  This enables to describe the 
flow on the energy.  When it moves inertially, both the energy and 
the momentum conservation laws are automatically involved in its 
transition amplitude.  This forbids the detector to respond in the 
Minkowski vacuum if $\Delta m\ge 0$.  On the other hand, if the 
detector moving in the classical potential $-Fz$ is considered, 
momentum conservation
in $z$ direction is not demanded because the translational 
invariance is broken.  Though energy conservation survives, it 
alone does not forbid the interaction mentioned above.  In this 
context, the recoil of the detector is essential as 
Parentani pointed out~\cite{pare}.  Decrease in the detector's 
kinetic energy is origin of energy paid for the radiation and 
increase in the detector's rest mass.  

The situation is rather similar to the case of rotating 
detectors though Bogoliubov coefficients between Minkowski and 
rotating modes have not been obtained~\cite{suga,ddm}.  
  The rotating detector can be excited 
and emit a particle in the Minkowski vacuum, energy of which is 
supplied via the recoil of the detector~\cite{suga}.  Circumstances 
would almost be the same even if other classical potentials are 
considered.  Therefore we can say the Unruh effect is not restricted 
to the case of uniformly acceleration  but is a rather general 
phenomenon in classical potential though distribution of detectors' 
energy gaps would not be like a thermal bath.  
\section*{Acknowledgments}

The authors thank Dr. K. Ikegami for stimulating 
discussions.

\end{document}